\documentclass{ieeeaccess}
\usepackage{cite}
\usepackage{amsmath,amssymb,amsfonts}
\usepackage{algorithm}
\usepackage{graphicx}
\usepackage{textcomp}
\usepackage{color}
\usepackage{algpseudocode}
\usepackage{subfigure}
\newtheorem{defin}{Definition}
\newtheorem{theorem}{Theorem}
\newtheorem{propo}{Proposition}
\newtheorem{example}{Example}

\def\BibTeX{{\rm B\kern-.05em{\sc i\kern-.025em b}\kern-.08em
    T\kern-.1667em\lower.7ex\hbox{E}\kern-.125emX}}

\begin{document}
\history{Date of publication xxxx 00, 0000, date of current version xxxx 00, 0000.}
\doi{10.1109/ACCESS.2021.DOI}

\title{Joint State Estimation Under Attack of Discrete Event Systems}
\author{\uppercase{Qi Zhang}\authorrefmark{1,2},
\uppercase{Carla Seatzu}\authorrefmark{2}, \IEEEmembership{Senior Member, IEEE}, \uppercase{Zhiwu Li}\authorrefmark{1,3}, \IEEEmembership{Fellow, IEEE}, \uppercase{and Alessandro Giua}\authorrefmark{2}, \IEEEmembership{Fellow, IEEE}}
\address[1]{School of Electro-Mechanical Engineering, Xidian University, Xi'an 710071, China}
\address[2]{Department of Electrical and Electronic Engineering, University of Cagliari, 09123 Cagliari, Italy}
\address[3]{Institute of Systems Engineering, Macau University of Science and Technology, Taipa 999078, Macau}
\tfootnote{This work was partially supported by the National Key R\&D Program of China under Grant 2018YFB1700104, the Natural Science Foundation of China under Grand Nos. 61472295, 61673309, 61873342, the ShaanXi Huashan Scholars, the Science and Technology Development Fund, MSAR, under Grant No. 122/2017/A3, and the Project RASSR05871 MOSIMA financed by Region Sardinia, FSC 2014-2020, annuity 2017, Subject area 3, Action Line 3.1.}

\markboth
{Q. Zhang \headeretal: Joint State Estimation Under Attack of Discrete Event Systems}
{Q. Zhang \headeretal: Joint State Estimation Under Attack of Discrete Event Systems}

\corresp{Corresponding author: Zhiwu Li (e-mail: zhwli@xidian.edu.cn).}

\begin{abstract}
The problem of state estimation in the setting of partially-observed discrete event systems subject to cyber attacks is considered. An operator observes a plant through a natural projection that hides the occurrence of certain events. The objective of the operator is that of estimating the current state of the system. The observation is corrupted by an attacker which can tamper with the readings of a set of sensors thus inserting some fake events or erasing some observations. The aim of the attacker is that of altering the state estimation of the operator. An automaton, called joint estimator, is defined to describe the set of all possible attacks. In more details, an unbounded joint estimator is obtained by concurrent composition of two state observers, the attacker observer and the operator observer. The joint estimator shows, for each possible corrupted observation, the joint state estimation, i.e., the set of states consistent with the uncorrupted observation and the set of states consistent with the corrupted observation. Such a structure can be used to establish if an attack function is harmful w.r.t. a misleading relation. Our approach is also extended to the case in which the attacker may insert at most $n$ events between two consecutive observations.
\end{abstract}


\titlepgskip=-15pt

\maketitle

\section{Introduction}\label{Introduction}

Cyber-physical systems are intelligent interconnected systems which are particularly exposed to network-based malicious attacks. Their security is a topic which during the last years has received much attention in different information and communications technology (ICT) communities such as automatic control \cite{DanZhang:2021,SurveyECC:2019}, computer science and engineering \cite{Uma:2013,Salahdine:2019}, and telecommunications \cite{Basit:2021}.

In the domain of automatic control, the security of dynamical systems has been addressed with two main formalisms. The first one is that of time-driven systems, either in continuous time \cite{XinHuang:2018,Rabehi:2019} or in discrete time \cite{HaoLiu:2021,WenleZhang:2021}. In \cite{XinHuang:2018}, the issue of reliable control in cyber-physical system under attack has been investigated. Rabehi \emph{et al.} \cite{Rabehi:2019} design a secure interval observer for solving the problem of state estimation. In \cite{HaoLiu:2021}, the issue of reachability analysis in discrete-time systems under attack has been studied. Finally, Zhang \emph{et al.} \cite{WenleZhang:2021} propose the problem of data-driven resilient control against cyber attacks.

The second formalism is that of discrete-event systems, where time driven dynamics are abstracted and a logical (non numerical) approach based on formal languages is adopted. The problem of attack detection in the framework of discrete event systems is addressed in  \cite{Thorsley:2006,Barbhuiya:2015,Fritz:2018}. In \cite{Agarwal:2019} the focus is on fault diagnosis of discrete event systems under attack. The problem of opacity enforcement by insertion functions under energy constraints has been investigated in \cite{Ji:2019}. The problem of supervisory control of discrete event systems under attack has been considered in \cite{Carvalho:2018,Lima:2017,Wakaiki:2019,LinAutomatica:2021,Goes:TAC,Zivana:2021,Xiang:CDC2020,Wang:2019a,DanTAC:2021,LinFeng:2021}.

Mainly inspired by some recent works \cite{Tong:2016,Goes:2020,Lima:2018,Su:2018}, we address the problem of state estimation in the setting of partially-observed discrete event systems subject to cyber attacks. In this paper, which is an extended version of \cite{Qi:2018}\footnote{In \cite{Qi:2018} we only provided preliminary ideas by considering a less general problem statement. Furthermore, no algorithm was formally presented therein.}, we consider a plant modeled as a discrete event system with state set $X$, whose evolution is observed by an operator. The occurrence of a subset of events $E_o$, called \emph{observable events}, can be detected by sensors while all other events, called \emph{silent events}, produce no observation. An evolution of the plant produces an observed word $s \in E_o^*$ which the operator uses  to  determine the set of \emph{consistent states} $\mathcal{C}(s) \subseteq X$, i.e., the set of states in which the system may be when $s$ has been produced.

We assume that  an attacker, which  has a full knowledge of the plant, may corrupt the sensor readings. This could happen because either the attacker can gain direct control of a sensor or it can corrupt messages between the plant and the operator assuming they are connected through a network. The particular attack model we adopt, among those that have been presented in the literature, is based on the one considered by Meira-G\'{o}es \emph{et al.} \cite{Goes:2020}. In particular, the attacker may insert in the word observed by the operator fake occurrences of \emph{compromised events} or, on the contrary, may erase the occurrence of such events. In addition, we consider the possibility that the length of a word inserted by the attacker between the occurrence of two observable events may be $n$-bounded.

The attacker aims to mislead an operator. Under attack, an observation $s \in E_o^*$ produced by a plant can be changed into a corrupted observation $s' \in E_o^*$ and as a result the operator computes a state estimate $\mathcal{C}(s') \subseteq X$, which is in general incorrect.  The attacker may have arbitrary goals. As an example, it may want to hide the fact that the plant has reached a critical state such that the operator does not activate appropriate protections that are fundamental for the system safeness. In all generality, we formalize a malicious goal by introducing a \emph{misleading relation} $\mathcal{R} \subseteq 2^X \times 2^X$  and we say that an attack is \emph{harmful} if there exists some observation $s$ that can be changed into a corrupted observation $s'$ such that $(\mathcal{C}(s),\mathcal{C}(s')) \in \mathcal{R}$.

In this paper, we first show how to construct two particular automata, called \emph{attacker observer} and \emph{operator observer}, which are defined on an augmented \emph{attack alphabet} which includes the observable events of the plant and the events describing the action of the attacker. Based on concurrent composition of these two observers, we design a \emph{joint estimator}, which describes all possible attacks. Finally, by inspection of such a structure one can determine if the attack function is harmful w.r.t. a misleading relation.

\subsection{Literature Review}

The problem of estimation under attack has been considered by relatively few authors in the automatic control literature.

In \cite{Ding:2017,Peng:2018,Cheng:2020} the state estimation problem for time-driven models is studied.

Ding \emph{et al.} \cite{Ding:2017} propose the problem of remote state estimation under denial-of-service attacks. A sensor needs to choose a channel to transmit the data packets, while an attacker needs to choose a channel to attack. They formalize such a problem as a two player stochastic game between the sensor and the attacker, and present optimal strategies in terms of computational complexity for both sides, respectively.

Peng \emph{et al.} \cite{Peng:2018} consider the issue of optimal attack energy allocation against remote state estimation. An optimal attack strategy that can result in maximal estimation error covariances is derived. Finally, they prove that the optimal strategy has a threshold structure.

Cheng \emph{et al.} \cite{Cheng:2020} develop an attack strategy that can degrade the estimation performance by tampering with the
sensors, eavesdropping the measurements, and injecting false feedback information. They conclude that, with the
presence of an attacker, the mean-squared stability condition of the state estimation is weakened.

In this work we consider discrete-event models, and the approaches developed in the above mentioned
literature cannot be adopted. The state estimation problem for discrete event systems so far has not been studied
in all generality, but has been partially addressed in the context of supervisory control. Here we mention a
few recent publications which have inspired our work.

Tong \emph{et al.} \cite{Tong:2016} present a new finite structure called parallel observer, which allows to simultaneously describe the observations of the supervisor and of the attacker. Based on the parallel observer, a maximally permissive supervisor is developed to enforce current-state opacity.

Meira-G\'{o}es \emph{et al.} \cite{Goes:2020} propose a novel bipartite transition structure in the framework of discrete event systems, namely, Insertion-Deletion Attack structure, and present a game-like relationship between the supervisor and the environment (the plant and the attacker) to determine if the attacker can lead the plant to a forbidden state without being detected by the supervisor.

Lima \emph{et al.}  \cite{Lima:2018} propose a defense policy that prevents cyber attacks at sensor and actuator layer in supervisory control systems. It is assumed that the attacker can alter the observation of events in a set of events $\Sigma_{vs}$, and modify the enabling of events in a set of events $\Sigma_{va}$. The detectable network attack security and undetectable network attack security are introduced to prevent the plant from reaching forbidden states.

Su \cite{Su:2018} addresses the problem of attack-with-bounded-sensor-reading-alteration (ABSRA), where the attacker can intercept the sensor readings from the plant and arbitrarily alter them but with an upper bound on the length of the altered observation word. In this way the attacker can cheat the supervisor, which will lead the plant to the undesirable states. The author also develops a supervisor that is robust to ABSRA.

We point out that significant differences exist between the problem setting considered in this paper and the problem setting dealt in most of the papers in the literature, including the above mentioned ones. What we propose is a methodology for studying how the possible choices of the attacker can affect the estimate of the operator. On the contrary, previous works on cyber attacks in the DES literature, including \cite{Carvalho:2018,Goes:2020,Lima:2018,Su:2018}, consider the case of an operator/supervisor: in such a case the goal of the attacker is to beguile the supervisor so that a specification is violated, i.e., the plant reaches a forbidden state or generates a forbidden evolution.

The proposed approach can be applied not only to the case in which an operator-supervisor controls a plant in closed-loop  --- as in the above mentioned papers --- but in more general settings, where the operator may have goals of different nature. As an example, in our framework one may study how cyber attacks disturb an operator-monitor which takes decisions based on its estimation of the plant state or an operator-diagnoser which aims to detect the occurrence of faults. \emph{Mutatis mutandis}, our approach can also be used for addressing a problem of opacity enforcing: in such a case the operator is an intruder that wants to infer a secret and the attacker is the agent that corrupts the observation to thwart the intruder.


\subsection{Contributions and structure of the paper}

This paper contains several original contributions.
\begin{itemize}
  \item The problem of \emph{joint state estimation under attack} is formalized; \emph{harmful attacks} are characterized in general terms by means of a \emph{misleading relation}.
  \item Based on the notion of \emph{attack alphabet} and \emph{attack words}, which describe how observations can be corrupted by the attacker we show how to construct two different observers: they describe the state estimates computed by the attacker and by the operator for each corrupted observation.
  \item A formal methodology to design a \emph{joint estimator} is presented. This automaton, constructed as the concurrent composition of the two observers, describes all possible attacks.
  \item The joint estimator shows, for each possible corrupted observation, the joint state estimation, i.e., the set of states consistent with the uncorrupted observation and the set of states consistent with the corrupted observation. Such a structure also indicates if a harmful attack exists.

\end{itemize}

The rest of the paper is organized as follows. In Section~\ref{Preliminaries}, we recall the basic notions of finite-state automata and of state estimation via observers. In Section~\ref{Problem}, we first describe the adopted attack model, and then formalize the problem considered in this paper. In Section~\ref{Attackeroperatorobservers}, we develop two observers: attacker observer and operator observer. In Section~\ref{Attackstructure}, we define the unbounded joint estimator as the concurrent composition of such observers. Then, we define an automaton that allows us, again via concurrent composition, to define a bounded joint estimator, starting from the unbounded one. Conclusions are finally drawn in Section~\ref{Conclusions} where we also discuss our future lines of research in this framework.

\section{Preliminaries}\label{Preliminaries}

Given an \emph{alphabet} $E$, let $E^*$ denote the set of all words on the alphabet. Given two words $w_1, w_2 \in E^*$, let $ w_1 w_2$ denote their \emph{concatenation}. Similarly, given two languages $L_1, L_2 \subseteq E^*$, we denote their concatenation $ L_1 L_2$ and when $L_1 = \{w \}$ we also write $ L_1 L_2 = w L_2$.

A \emph{deterministic finite-state automaton} (DFA) is a four-tuple $G=(X,E,\delta,x_0)$, where \emph{X} is the set of states, \emph{E} is the set of events (alphabet), $\delta:X\times E\rightarrow X$ is the transition function, and $x_0$ is the initial state. The transition function can be extended to $\delta^*:X\times E^*\rightarrow X$ such that $\delta^* (x,\varepsilon)=x$, and $\delta ^*(x,\sigma e)=\delta(\delta^* (x,\sigma),e)$ for all $x \in X$, $e \in E$ and $\sigma \in E^*$. The \emph{generated language} of \emph{G} is defined as $L(G)=\{\sigma\in E^* \ | \ \delta^* (x_0,\sigma) \text { is defined}\}$.

Given two alphabets $E'$ and $E$ with $E' \subseteq E$, the \emph{natural projection} on $E'$, $P_{E'}:E^*\rightarrow (E')^*$ is defined as \cite{Wonham:1989}:
\begin{equation}\label{naturalprojection}
P_{E'}(\varepsilon):=\varepsilon \ , \
P_{E'}(\sigma e) := \left\{
             \begin{array}{lcl}
             {P_{E'}(\sigma) e} \ &\text{if}& \ e \in E', \\
             {P_{E'}(\sigma)}   \ &\text{if}& \ e \in E \setminus E'.
             \end{array}
        \right.
\end{equation}
Therefore, given a word $\sigma \in E^*$, its natural projection on $E'$ is obtained by erasing events that do not belong to $E'$.

The \emph{concurrent composition} of two languages is defined as $L_1 \parallel L_2=\{\sigma \in E^*\ | \ P_{E_1}(\sigma) \in L_1, P_{E_2}(\sigma) \in L_2\}$, where $E_1$ and $E_2$ are alphabets of $L_1$ and $L_2$, respectively, and $E=E_1 \cup E_2$.
The concurrent composition operator can also be defined for DFA. In particular, given two DFA $G'$ and $G''$, their {\em concurrent composition}, denoted as $G=G' \parallel G''$, generates language $L(G)=L(G') \parallel L(G'')$.

A \emph{partially-observed deterministic finite-state automaton} is denoted as $G=(X,E,\delta,x_0)$, where $E=E_o\cup E_{uo}$, $E_o$ is the \emph{set of observable events}, and $E_{uo}$ is the \emph{set of unobservable events}. In the following, to keep the notation simple, we denote as $P:E^*\rightarrow E_o^*$ the natural projection on $E_o$. The \emph{inverse projection} $P^{-1}:E_o^* \rightarrow 2^{E^*}$ is defined as $P^{-1}(s)=\{\sigma \in E^*:P(\sigma)=s\}$, where $\sigma \in E^*$, and $s \in E_o^*$.

When a partially-observed DFA generates a word $\sigma \in L(G)$ it produces the observation $s = P(\sigma) \in E_o^*$. However, in general a given observation may be produced by more than one generated word. The \emph{set of words consistent with observation $s$} is defined by
$$
\mathcal{S}(s) = P^{-1}(s) \cap L(G) = \{ \sigma \in L(G) \mid P(\sigma) = s \}
$$
and denotes the set of words generated by the DFA that produce observation $s$.

Correspondingly, the \emph{set of states consistent with observation $s$} is defined by
$$
\mathcal{C}(s) =  \{ x \in X  \mid  (\exists \sigma \in \mathcal{S}(s)) \ \delta^*(x_0,\sigma) = x \}
$$
and denotes the set of states in which the DFA can be when observation $s$ had been produced.

The set $\mathcal{C}(s) \subseteq  X $ is also called the \emph{state estimate} corresponding to observation $s \in E_o^*$. The problem of state estimation of a partially-observed DFA $G$ can be solved, in all generality, constructing a new structure called its \emph{observer} (see \cite{Lafortune:2008} for details).

Let us first define, for a partially-observed DFA $G$, the \emph{unobservable reach} of a state $x\in X$. This set is denoted by $UR(x)$ and is defined as a set of states $x' \in X$ reached from state $x$ generating an unobservable word $\sigma \in E_{uo}^*$, i.e., $UR(x) =\{x' \ | \ \exists \sigma\in E_{uo}^*, \ \delta^*(x,\sigma)=x'\}$. This definition can be extended to a set of states $B \subseteq 2^X$ as follows:
$$UR(B)=\bigcup \limits_{x\in B}UR(x).$$

The \emph{observer} $Obs(G)$ of a partially-observed DFA $G=(X,E,\delta,x_0)$, is a DFA:
$$Obs(G) = (B,E_o,{\delta _{obs}},b_0),$$ where the set of states is $B\subseteq 2^X$, the alphabet $E_o$ is the set of observable events of $G$, the transition function $\delta _{obs}: B\times E_o \rightarrow B$ is  defined as:
$$\delta _{obs}(b,e_o):=\bigcup \limits_{x\in b}UR(\{ x' \mid \delta(x,e_o)=x'\}),$$
 and the initial state is $b_0:=UR(x_0)$.

As shown in \cite{Lafortune:2008}, given a partially-observed DFA $G$ with observer $Obs(G) = (B,E_o,{\delta _{obs}},b_0)$, for any observation $s \in E_o^*$ produced by $G$ it holds that
$\mathcal{C}(s) = \delta _{obs}^*(b_0,s)$.

\section{The joint state estimation problem}\label{Problem}

In this section, first, we introduce an attack model and an original formalism to represent it. Then, we formalize the problem considered in this paper.

\subsection{Attack model}\label{AttackModel}

\begin{figure}[htbp]
  \centering
  \includegraphics[width=2.5in]{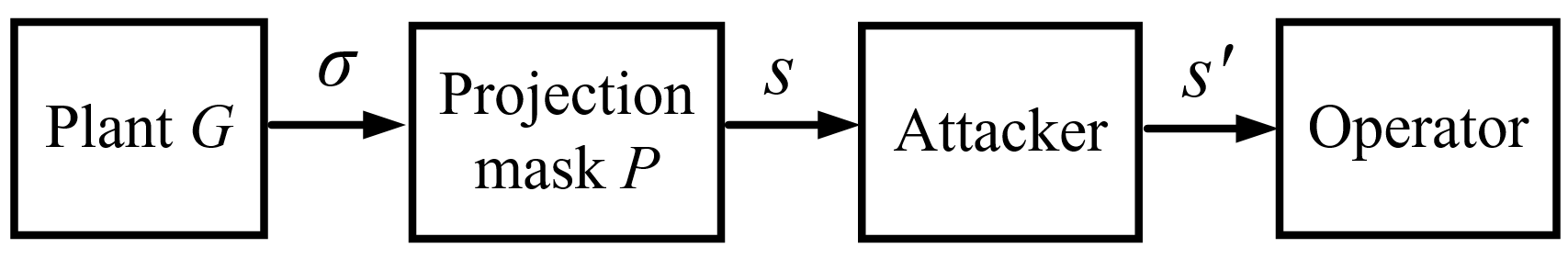}\\
  \caption{A plant $G$ under attack.}
  \label{fig3}
\end{figure}

In this paper we consider a plant modeled by a partially observable DFA with set of observable events $E_o$ and set of unobservable events $E_{uo}$. Referring to Fig.~\ref{fig3}, if $\sigma$ is a word generated by the plant, the \emph{observed word} is $s=P(\sigma)$. An attacker may corrupt the output signals produced by the plant with the effect of inserting in the observation some events that did not occur, or erasing some events that have occurred. Such a \emph{corrupted observation} is denoted as $s'$ (a sequence of events in $E_o$), and the operator constructs its \emph{state estimation} based on $s'$. In our framework, we assume the operator monitors the plant to estimate its current state: the objective of the attacker is to corrupt the observation in such a way that a correct estimation is not possible.

\begin{defin} \cite{Goes:2020} The \emph{set of compromised events} is denoted as ${E_{com}} \subseteq {E_o}$. It includes all the observable events that can be corrupted by the attacker, either inserting them in the operator observation, even if they have not actually occurred, or erasing them in the operator observation.~\hfill $\diamond$
\label{def2}
\end{defin}

The definition of compromised events was first proposed in \cite{Goes:2020}. However, while in \cite{Goes:2020} the authors assume that all the compromised events can be inserted and erased by the attacker, here we slightly generalize the definition as follows.

The set of compromised events that can be inserted in the observer evolution is denoted as $E_{ins}$, and the set of events that can be erased is denoted as $E_{era}$. To keep the presentation general, we assume that $E_{ins}$ and $E_{era}$ are not necessarily disjoint.

\begin{figure}[htbp]
  \centering
  \includegraphics[width=1.3in]{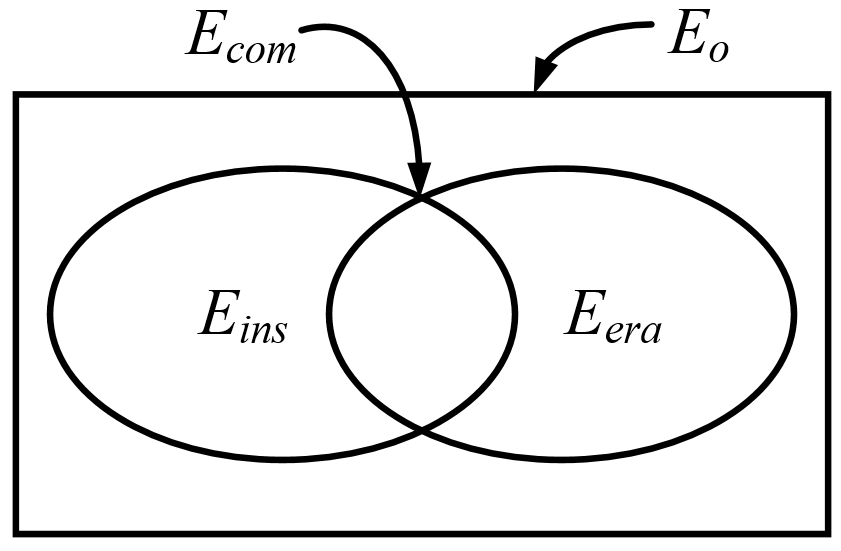}\\
  \caption{The relationship among the subsets of $E_o$.}
  \label{fig4}
\end{figure}

The relationship among the different subsets of observable events $E_o$ is clarified in Fig.~\ref{fig4}.

We now formally describe the action of the attacker in terms of two new types of events that it can generate. More precisely, even if it is possible to directly define the attacker as a finite-state transducer that ``translates" an observed word $s$ into a corrupted observation $s'$ (see Fig.~\ref{fig3}), for a reason that will appear clear in the following, we prefer to characterize the attacker's action in terms of a new word defined on a so-called \emph{attack alphabet} $E_a$.

\begin{defin} The \emph{attack alphabet} is defined as $E_a=E_o \cup E_+ \cup E_-$, and we assume that $E_o$, ${E_+}$, and ${E_-}$ are disjoint sets.
\label{def3}
\end{defin}

The \emph{set of inserted events} \cite{Goes:2020} is denoted as ${E_+}$, namely $E_+=\{e_+ \mid e \in E_{ins}\}$. The occurrence of an event $e_+ \in E_+$ denotes the fact that the attacker inserts in the operator observation an event $e \in E_{ins}$ that has not occurred in reality.

The \emph{set of erased events} \cite{Goes:2020} is denoted as ${E_-}$, namely $E_-=\{e_- \mid e \in E_{era}\}$. The occurrence of an event $e_- \in E_-$ denotes the fact that the attacker erases from the operator's observation event $e \in E_{era}$ generated by plant.~\hfill $\diamond$

Given a bound $n \in \mathbb{N} \cup \{ \infty \}$, let $E^{\leq n}_+=\{w_+ \in E ^*_+ \ \mid \ |w_+| \leq n\}$ denote the set of words on alphabet $E_+$ whose length does not exceed $n$. Note that if $n=\infty$ then $E^{\leq n}_+ = E ^*_+$.

\begin{defin} Given a plant $G$ with a set of compromised events $E_{com}=E_{ins} \cup E_{era}$, let $n \in \mathbb{N} \cup \{ \infty \}$ be a bound. An $n$-bounded attacker can be defined by an \emph{attack function} $f_n: P(L(G)) \rightarrow E_a^\ast$, where $E_a$ is the attack alphabet (Definition~\ref{def3}), satisfying the following conditions:
\label{def4}
\end{defin}

\begin{enumerate}
\item[(a)] $f_n(\varepsilon)\in E_+^{\leq n}$,

\item[(b)] $\forall se \in P(L(G))$ with $s \in E_o^*$:
\begin{equation}\label{equationb}
\left\{ \begin{array}{ll} f_n(se) \in f_n(s)\{e_-,e\}E_+^{\leq n} &\text{if} \quad e \in E_{era}, \\  f_n(se) \in f_n(s) \{ e \} E_+^{\leq n} &\text{if} \quad e \in E_o \setminus E_{era}.   \end{array} \right.
\end{equation}
\end{enumerate}~\hfill $\diamond$

In Definition~\ref{def4}, condition (a) means that the attacker can insert a bounded word $w_+ \in E_+^{\leq n}$ at the initial state, before any event generated by the plant is observed. {Condition (b) implies that if an event $e \in E_{era}$ occurs, the attacker can either erase event $e$ or not erase it, and then insert any word $w_+ \in E_+^{\leq n}$. If an event $e \in E_o \backslash E_{era}$ occurs, then the attacker can insert any word $w_+ \in E_+^{\leq n}$ after $e$.}

We notice that imposing that the attacker may insert at most $n$ consecutive events between any two observed events, makes sense in practice. Indeed, even if our model is purely logical, a real system can produce in a finite time just a finite number of events. If the attacker could introduce an arbitrarily large number of events between two consecutive observed events, this would lead to an anomalous behavior as observed by the operator.

We denote as ${\mathcal F}_n$ the set of attack functions for a given $n\in \mathbb{N} \cup \{\infty\}$.

\begin{defin}
The language modified by an attack function $f_n$ is called {\em attack language}. It is denoted as $L(f_n,G)$ and is defined as $L(f_n,G)=f_n(P(L(G)))$. A word $w \in L(f_n,G)$ is called an \emph{attack word}.

The set of all the attack languages relative to a given $n\in \mathbb{N} \cup \{\infty\}$, denoted as $L({\mathcal F}_n,G)$, is defined as
\begin{equation}\label{attackfunctions}
L({\mathcal F}_n,G)=\bigcup_{f_n\in {\mathcal F}_n}L(f_n,G)=\bigcup_{f_n\in {\mathcal F}_n}f_n(P(L(G))).
\end{equation}~\hfill $\diamond$
\label{defFn}
\end{defin}

Given two integer numbers $n$ and $n'$, ${\mathcal F}_n \subseteq {\mathcal F}_{n'}$ if {$n \leq n'$.} Furthermore, ${\mathcal F}_n\subseteq {\mathcal F}_{\infty}$ for all $n<\infty$.

\begin{defin}
The \emph{operator mask} $\widehat{P}:E_a^* \rightarrow E_o^*$ is defined as:
\label{def5}
\end{defin}
\begin{equation}\label{reductionprojection}
\widehat{P}(\varepsilon)=\varepsilon, \
\widehat{P}(w e')=\left\{
             \begin{array}{lcl}
             {\widehat{P}(w)e} \text{ if } e'=e \in E_o \ \vee \\
             \qquad \quad \quad e'=e_+ \in E_+, \\
             {\widehat{P}(w)} \ \text{ if } e'=e_- \in E_-.
             \end{array}
        \right.
\end{equation}~\hfill $\diamond$

\begin{figure}[htbp]
  \centering
  \includegraphics[width=2.3in]{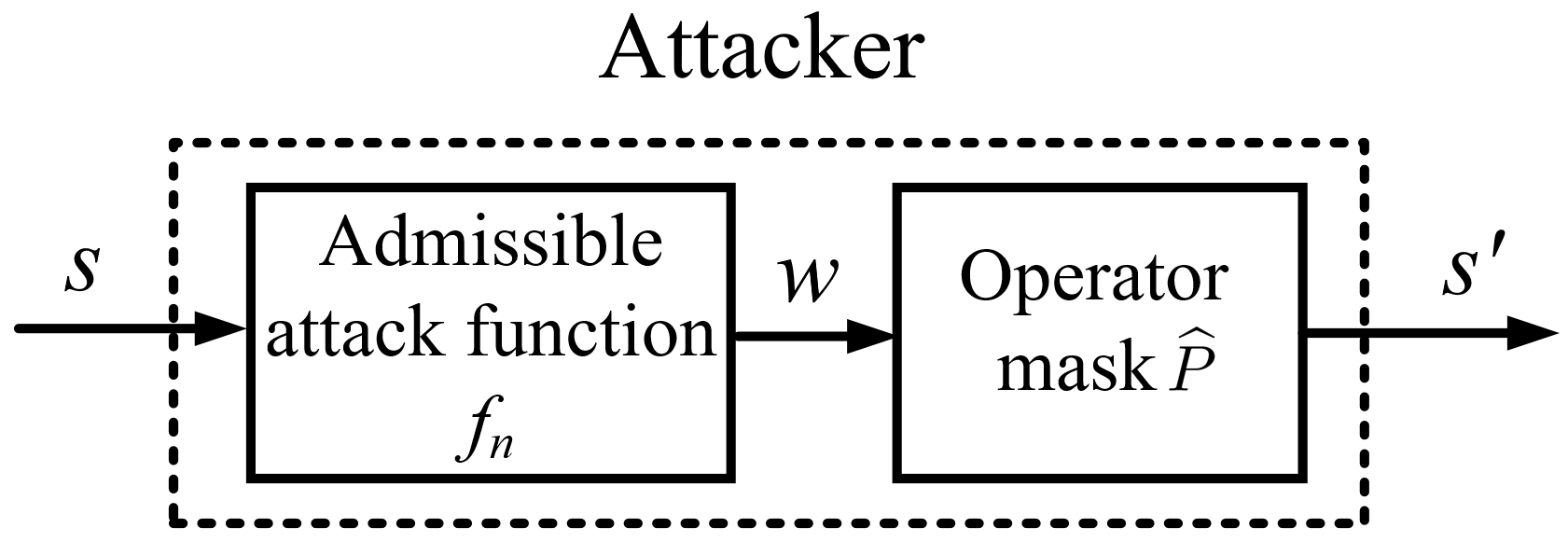}\\
  \caption{Internal structure of the attacker with observed word $s \in E_o^*$, attack word $w \in E_a^*$, and corrupted observation $s' \in E_o^*$.}
  \label{fig5}
\end{figure}

The internal structure of the attacker is visualized in Fig.~\ref{fig3} as a black box taking an observation $s$ as an input and producing a corrupted observation $s'$ as an output. Such an internal structure is sketched in more detail in Fig.~\ref{fig5}.

Here the observed word is $s=P(\sigma)$ (a sequence of events in $E_o$). The attacker corrupts the observation according to the attack function $f_n$, producing $w\in L(f_n,G) \subseteq E_a^*$. Such a sequence is projected via $\widehat P$ on $E_o$, generating a word $s'$. The plant operator constructs its state estimation based on $s'$.

\subsection{Problem statement}\label{Statement}

In this subsection we first describe the possible goals of the attacker on which we are focusing in this paper. Then we formalize the problem statement. To this aim we introduce a relation $\mathcal{R} \subseteq 2^X \times 2^X$, called a \emph{misleading relation}. If $s$ and $s'$ in $E_o^*$ denote, respectively, the generic uncorrupted and corrupted observation of the operator, the goal of the attacker is achieved whenever
$$
(\mathcal{C}(s),\mathcal{C}(s'))\in \mathcal{R},
$$
i.e., whenever the pair (set of states consistent with the uncorrupted observation, set of states consistent with the corrupted observation) belongs to $\mathcal{R}$. The following definition formalizes this.

\begin{defin}
Let $G=(X,E,\delta,x_0)$ be a plant with set of observable events $E_o$ and its observer $Obs(G)=(B,E_o,\delta_{obs},b_0)$. An attack function $f_n$ is \emph{harmful} w.r.t. a relation $\mathcal{R} \subseteq 2^X \times 2^X$ if there exists an observation $s \in P(L(G))$ generated by the plant, whose set of consistent states is $\mathcal{C}(s)= \delta_{obs}^*(b_0,s)$, such that $s$ can be corrupted into a word $s' = \widehat{P}(f_n(s))$ whose corresponding set of consistent states is $\mathcal{C}(s') = \delta_{obs}^*(b_0,s')$, and $(\mathcal{C}(s),\mathcal{C}(s')) \in \mathcal{R}$.~\hfill $\diamond$
\label{def7}
\end{defin}

Different physical problems can be described in this setting by suitably defining relation $\mathcal{R}$. A significant example is the following one.

\begin{example}
Assume that a subset of states of a system, $X_{cr} \subseteq X$, is labeled as critical in the sense that when the system is in one of such states, a protective action must be taken to avoid possible damages. The operator is monitoring the system evolution in order to establish when one of such states is reached. Obviously, if the operator does not realize that a critical state has been reached, no action is taken and the system behaviour can be seriously compromised. The effectiveness of an attacker that aims to affect the system observation in order to prevent the operator to realize when a critical state is reached, can be evaluated via the misleading relation defined as $\mathcal{R} = \{ (X',X'') \ | \ X' \cap X_{cr} \neq \emptyset \text { and } X'' \cap X_{cr} = \emptyset \}$.  Indeed, the attack is harmful if there exists at least one observation $s$ such that $\mathcal{C}(s) \cap X_{cr} \neq \emptyset$ (meaning that the system may be in a critical state) which can be corrupted to an observation $s'$ with $\mathcal{C}(s') \cap X_{cr} = \emptyset$ (meaning that the operator excludes the possibility that the system is in a critical state).~\hfill $\diamond$
\end{example} \label{ExaProblem}

Given a plant $G=(X,E,\delta,x_0)$ with set of observable events $E_o$, and a misleading relation $\mathcal{R} \subseteq 2^X \times 2^X$, the main contribution of this paper consists in providing a tool, called a joint estimator that contains all the possible actions (insert and/or erase observations) that an attacker may implement during the system evolution. On the basis of such a structure, the attacker can establish if the attacks are harmful to the plant.

\section{Attacker Observer and Operator Observer}\label{Attackeroperatorobservers}

In this section we introduce two special structures, called {\em Attacker Observer} and {\em Operator Observer}, which are fundamental to derive the solution to the joint estimation problem we are considering.

\subsection{Attacker Observer}\label{Attackerobserver}

The attacker observer $Obs_{att}(G)$ describes all possible attack words that can be generated by functions in $\mathcal{F}_{\infty}$ and the corresponding sets of consistent states of the system. Since attacks are performed by the attacker, it knows which observations originate from events that have really occurred in the plant ($E_o$), which observations have been erased ($E_-$), and which observations have been inserted ($E_+$). The attacker observer $Obs_{att}(G)$ can be constructed using Algorithm~\ref{alg1}.

\begin{algorithm}[htbp]
\caption{Construction of the attacker observer $Obs_{att}(G)$}
\label{alg1}
\begin{algorithmic}[1]
\Require
An observer $Obs(G) = (B,{E_o},{\delta _{obs}},b_0)$, $E_{ins}$, and $E_{era}$.
\Ensure
An attacker observer $Obs_{att}(G) = \left(B,E_a,\delta _{att},b_0 \right)$.
\State Let ${E_a} := E_o \cup {E_+} \cup {E_-}$;
\State Let $\delta_{att}:=\delta_{obs}$;
\ForAll{$e\in E_{era}$,}
\ForAll{$b\in B$,}
\If{$\delta_{att}(b,e)=b',$}
\State $\delta_{att}(b,e_-)=b'$;
\EndIf
\EndFor
\EndFor
\ForAll{$e\in E_{ins}$,}
\ForAll{$b\in B$,}
\State $\delta_{att}(b,e_+)=b$;
\EndFor
\EndFor
\end{algorithmic}
\end{algorithm}

According to Algorithm~\ref{alg1}, the set $E_a$ is initially computed and the transition function of $Obs_{att}(G)$ is initialized at $\delta_{att}=\delta_{obs}$. Indeed, events in $E_o$ are events actually occurring in the plant, thus when such events occur the attacker updates its state estimation according to the transition function of $Obs(G)$.

Then, for all $e \in E_{era}$ and for all $b\in B$, whenever $\delta_{att}(b,e)$ is defined, the algorithm imposes $\delta_{att}(b,e_-)=\delta_{att}(b,e)$. Indeed, the attacker knows that $e_-$ corresponds to event $e$ that has been canceled, thus the way it updates its estimation is the same in the case of $e$ and $e_-$.

Finally, for all events $e\in E_{ins}$, and for all states $b\in B$, we add self-loops $\delta_{att}(b,e_+)=b$. Indeed, the attacker knows that events in $E_+$ are fake events that have not really occurred in the plant, thus it does not update its estimation based on them. In particular, self-loops correspond to the possibility of inserting an arbitrarily large number of such events, which is consistent with the fact that we are dealing with attack functions in ${\mathcal F}_{\infty}$.

\begin{figure}[htbp]
  \centering
  \subfigure[$G$]{
    \includegraphics[width=2.4in]{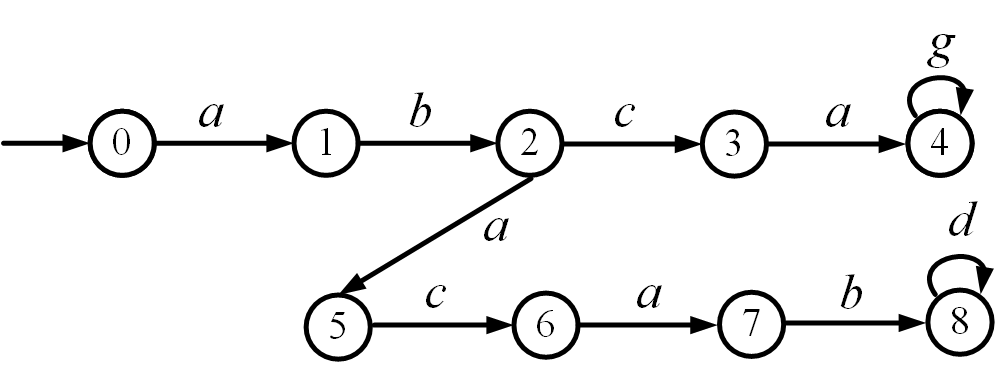}
  }
  \subfigure[$Obs(G)$]{
    \includegraphics[width=2.4in]{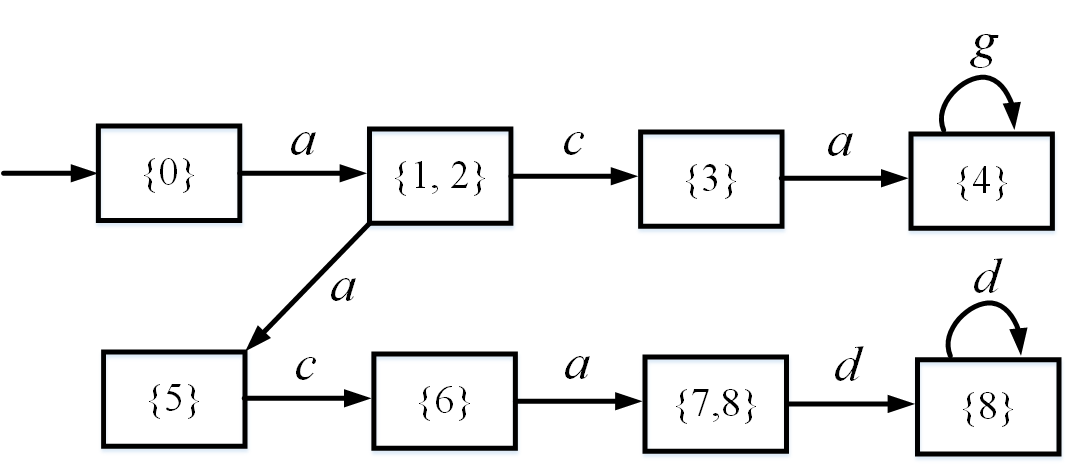}
  }
  \caption{(a) A partially-observed plant $G$; (b) its observer $Obs(G)$, where $E_o=\{a,c,d,g\}$.}
  \label{fig2} 
\end{figure}

\begin{figure}[htbp]
  \centering
  \subfigure[$Obs_{att}(G)$]{
    \includegraphics[width=2.5in]{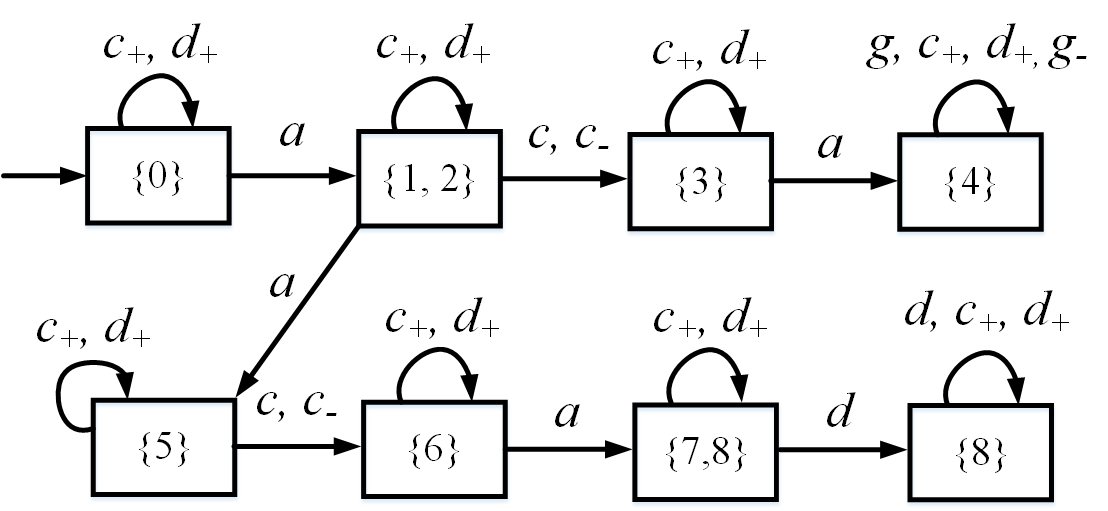}
  }

  \subfigure[$Obs_{opr}(G)$]{
    \includegraphics[width=3in]{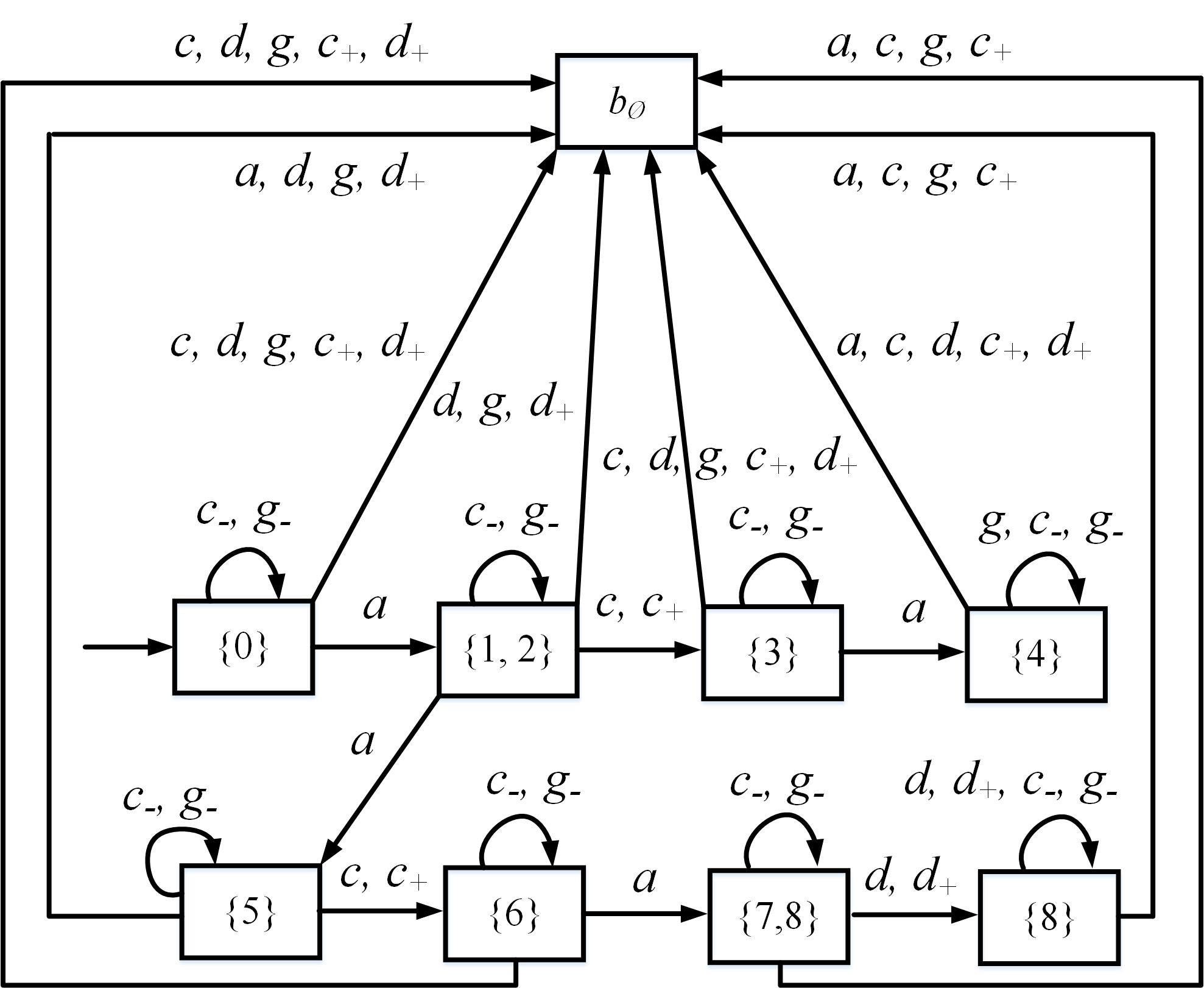}
  }

  \caption{(a) Attacker observer in Example \ref{exa3} and (b) operator observer in Example \ref{exa5} for the plant in Fig.~\ref{fig2}.}

  \label{fig7} 
\end{figure}

\begin{example} Consider a partially-observed plant $G = \left( {X,E,\delta ,{x_0}} \right)$ in Fig.~\ref{fig2}(a), where $E = {E_o} \cup {E_{uo}}$, $E_o=\{a,c,d,g\}$, and $E_{uo} = \left\{ b \right\}$. The corresponding observer of \emph{G} is shown in Fig.~\ref{fig2}(b). Let $E_{ins}=\{c,d\}$, and $E_{era}=\{c,g\}$. The attacker observer constructed using Algorithm~\ref{alg1} is shown in Fig.~\ref{fig7}(a). \label{exa3} \end{example}

Since events $c,g \in E_{era}$, and there is a transition labeled $c$ from state $\{1, 2\}$ to state $\{3\}$ in the observer of the plant $Obs(G)$, we add transitions labeled $c$ and $c_-$ from state $\{1, 2\}$ to state $\{3\}$ in the attacker observer. Similar arguments can be used to explain transitions labeled $c$ and $c_-$ from state $\{5\}$ to state $\{6\}$, self-loops labeled $g$ and $g_-$ at state $\{4\}$. Then, since $c$, $d \in E_{ins}$, we add self-loops labeled $c_+$ and $d_+$ at all the states.~\hfill $\diamond$

The following proposition provides a characterization of the attacker observer.

\begin{propo} Consider a plant $G$ with set of observable events $E_o$ and observer $Obs(G) = (B,{E_o},{\delta _{obs}},b_0)$.
Let $f_\infty$ be an attack function, $\mathcal{F}_\infty$ be the set of attack functions, and $E_a = E_o \cup E_+ \cup E_-$ be an attack alphabet. Let $Obs_{att}(G)$ be the attacker observer constructed using Algorithm~\ref{alg1}. It holds that:
\label{prop1} \end{propo}

\begin{enumerate}
\item[(a)] $L(Obs_{att}(G))=L({\mathcal F}_{\infty},G)$;

\item[(b)] $\forall s \in P(L(G))$, $\forall f_{\infty}\in {\mathcal F}_{\infty}$ with $w=f_\infty(s) \in E_a^*$: $$\delta_{att}^*(b_0,w)=\delta_{obs}^*(b_0,s).$$
\end{enumerate}

\emph{Proof:} (a) According to Algorithm~\ref{alg1}, Step~2 implies that $L(Obs_{att}(G))$ contains all words that can be observed if no attack occurs. Correspondingly, according to the definition of attack function (Definition~\ref{def4}), the set of attack languages $L({\mathcal F}_{\infty},G)$ also contains these words since the attacker does not reduce the language of the plant.

Steps~3--9 guarantee that all attacks resulting from the cancellation of events in $E_{era}$ are considered. Correspondingly, according to the definition of attack function, each time the plant generates an event $e \in E_{era}$, then the attacker can erase it.

Finally, Steps~10--14 guarantee that all attacks resulting from the insertion of an arbitrarily large number of events in $E_{ins}$ are taken into account. Again, according to the definition of attack function, the attacker can insert any word $w_+ \in E_+^{\leq n}$ whenever possible.

Thus, we can conclude that $L(Obs_{att}(G))=L({\mathcal F}_{\infty},G)$.

(b) We prove this by induction on the length of $s$. If $s=\varepsilon$, the result follows from the fact that, by definition of attack function, it is $f_{\infty}(\varepsilon)\in E_+^{\leq n} $, and by Steps~10--14, events in $E_+$ lead to self-loops in $Obs_{att}(G)$.

Let us now consider a generic word $s \in P(L(G))$ with length greater than one, written as $s=\overline{s} e$, where $\overline{s} \in P(L(G))$ and $e\in E_o$. Assume the result holds for $\overline{s}$. We prove that it also holds for $s=\overline{s} e$ considering the following two possible cases.

If $e\in E_{era}$, by the definition of attack function, $w\in  \bigcup \limits_{f_\infty \in {\mathcal F}_\infty } f_{\infty}(\overline{s})\{e_-,e\}E_+^{\leq n}$ is true. According to Steps~3--9, events $e$ and $e_-$ are dealt with in the same manner when defining the transition function $\delta_{att}$. Finally, as just pointed out, according to Steps~10--14, events in $E_+$ lead to self-loops in $Obs_{att}(G)$.

Finally, if $e\in E_o \setminus E_{era}$, by the definition of attack function, $w \in  \bigcup \limits_{f_\infty \in {\mathcal F}_\infty } f_{\infty}(\overline{s}) e E_+^{\leq n}$ is true. Thus the result follows from the fact that, according to Steps~10--14, events in $E_+$ lead to self-loops in $Obs_{att}(G)$ and events in $E_o$ are dealt with in the same manner in $Obs_{att}(G)$ and $Obs(G)$.~\hfill $\Box$

Now we discuss the computational complexity of constructing the attacker observer $Obs_{att}(G)$. Given a plant \emph{G} with set of states \emph{X}, the observer of the plant $Obs(G)$ can be constructed in $2^{|X|}$ steps. According to Algorithm~\ref{alg1}, $Obs_{att}(G)$ has the same number of states as $Obs(G)$; thus the complexity of building $Obs_{att}(G)$ is $O(2^{|X|})$.

\subsection{Operator Observer}\label{Operatorobserver}

The attack model we are considering may change an observation $s$ into a corrupted observation $s'$ which cannot be produced by the nominal plant. In this case the operator understands that the system is under attack. This can be formalized as follows.

\begin{defin} Consider a plant $G$. An attack function $f_n$ is said to be \emph{stealthy} if $\widehat{P}(L(f_n,G)) \subseteq P(L(G))$, where $\widehat{P}$ is the operator mask.~\hfill $\diamond$
\label{def6}
\end{defin}

In words, stealthiness requires that the set of words that an operator observes when the system is under attack is contained in the set of words the operator may observe when no attack occurs.

The operator observer $Obs_{opr}(G)$ generates two different sets of words. The first set includes all words on $E_a^*$ that may either result from an uncorrupted observation of the plant or from a corrupted observation which keeps the attacker stealthy. The second set of words includes all the previous words continued with a symbol in $E_a$ so that the resulting word is not consistent with an uncorrupted observation. While the words in the first set lead to a set of states that according to the operator are consistent with the perceived observation, those in the second set lead to a dummy state denoted as $b_\emptyset$. The operator observer $Obs_{opr}(G)$ can be constructed using Algorithm~\ref{alg2}, as shown below.

\begin{algorithm}[htbp]

\caption{Construction of the operator observer $Obs_{opr}(G)$}
\label{alg2}
\begin{algorithmic}[1]
\Require
An observer $Obs(G) = (B,{E_o},{\delta _{obs}},b_0)$, $E_{ins}$, and $E_{era}$.
\Ensure
An operator observer $Obs_{opr}(G) = \left(B_{opr},E_a,\delta_{opr},b_0\right)$.
\State Let $B_{opr}:=B \cup \{b_\emptyset\}$;
\State Let $E_a:= E_o \cup E_+ \cup E_-$;
\State Let $\delta_{opr}:=\delta_{obs}$;
\ForAll{$e\in E_{ins}$,}
\ForAll{$b\in B$,}
\If{$\delta_{opr}(b,e)=b'$,}
\State $\delta_{opr}(b,e_+)=b'$;
\EndIf
\EndFor
\EndFor
\ForAll{$e\in E_{era}$,}
\ForAll{$b\in B$,}
\State $\delta_{opr}(b,e_-)=b$;
\EndFor
\EndFor
\ForAll{$e_a \in E_a$,}
\ForAll{$b\in B$,}
\If{$\delta_{opr}(b,e_a)$ is not defined,}
\State $\delta_{opr}(b,e_a)=b_\emptyset$;
\EndIf
\EndFor
\EndFor
\end{algorithmic}
\end{algorithm}

According to Algorithm~\ref{alg2}, the set of states $B_{opr}=B\cup \{b_{\emptyset}\}$ and the set of events $E_a$ are initially computed. Then, the transition function of $Obs_{opr}(G)$ is initialized at $\delta_{opr}=\delta_{obs}$. Indeed, events in $E_o$ are events actually occurring on the plant; when such events occur, the operator updates its state estimation according to the transition function of $Obs(G)$.

Furthermore, for all $e\in E_{ins}$ and for all $b\in B$, we impose $\delta_{opr}(b,e_+)=\delta_{opr}(b,e)$. Indeed, the operator does not distinguish between events in $E_+$ and the corresponding events in $E_{ins}$. For all $e \in E_{era}$ and for all $b\in B$, we add self-loops $\delta_{opr}(b,e_-)=b$. Indeed, events in $E_-$ correspond to no observation by the operator.

Finally, for all the events $e_a\in E_a$ that are not enabled at the generic state $b\in B$, let $\delta_{opr}(b,e_a)=b_\emptyset$. As a result, for all $b\in B$ and for all $e_a\in E_a$, function $\delta_{opr}(b,e_a)$ is defined. On the contrary, $\delta_{opr}(b_\emptyset,e_a)$ is undefined for all $e_a\in E_a$.

\begin{example} Consider again the plant $G$ in Fig.~\ref{fig2}. Let $E_{ins}=\{c,d\}$ and $E_{era}=\{c,g\}$. The operator observer constructed using Algorithm~\ref{alg2} is visualized in Fig.~\ref{fig7}(b).
\label{exa5} \end{example}

Since $c,d \in E_{ins}$ and there is a transition labeled $c$ from state $\{1, 2\}$ to state $\{3\}$ in $Obs(G)$, we add transitions labeled $c$ and $c_+$ from state $\{1, 2\}$ to state $\{3\}$ in the operator observer. Similar arguments can be used to explain the transitions labeled $c$ and $c_+$ from state $\{5\}$ to state $\{6\}$, the transitions labeled $d$ and $d_+$ from state $\{7,8\}$ to state $\{8\}$, and the self-loops labeled $d$ and $d_+$ at state $\{8\}$. Then, since $c$, $g \in E_{era}$, we add self-loops labeled $c_-$ and $g_-$ at all the states. Finally, we add all the missing transitions to the new state $b_\emptyset$, which has no output arc.~\hfill $\diamond$

To better characterize the properties of the operator observer, let us define two sets of words associated to an attack model as described in Subsection~\ref{AttackModel}.

\begin{defin} The \emph{set of stealthy words} on  attack alphabet $E_a$ is
$$
W_s=\{w \in E_a^* \ | \ \widehat{P}(w) \in P(L(G))\},
$$
while  the \emph{set of exposing words} on $E_a$ is
$$
W_e = \{ w e_a \in E_a^*  \mid w \in W_s, e_a \in E_a, w e_a \not\in W_s \}.
$$
~\hfill $\diamond$
\label{def_stex}
\end{defin}

In plain words, a stealthy word $w$ produces an observed word $s' = \widehat{P}(w)$ which does not  reveal the presence of the attacker because such an observation may have been produced by the plant. An exposing word is a non-stealthy word on $E_a$ whose strict prefixes are all stealthy: hence it reveals the presence of the attacker but only at the last step.

The following proposition provides a characterization of the operator observer.

\begin{propo} \label{prop2} Let $G$ be a plant with set of observable events $E_o$ and observer $Obs(G) = (B,{E_o},{\delta _{obs}},b_0)$. Given an attack alphabet $E_a = E_o \cup E_+ \cup E_-$ with set of stealthy words $W_s$ and set of exposing words $W_e$, let $Obs_{opr}(G)$ be the operator observer constructed by Algorithm~\ref{alg2}. It holds that:
\end{propo}

\begin{enumerate}
\item[(a)] $L(Obs_{opr}(G)) = W_s \cup W_e$;

\item[(b)] $\forall w \in L(Obs_{opr}(G))$: if $w \in W_s$, then $\delta_{opr}^*(b_0,w)=\delta_{obs}^*(b_0,\widehat{P}(w))$; if $w \in W_e$, then $\delta_{opr}^*(b_0,w)=b_{\emptyset}$.
\end{enumerate}

\emph{Proof:} (a) Follows from Algorithm~\ref{alg2}, and from the definitions of stealthy words, exposing words, and operator mask.

In more detail, Step~3 guarantees that all uncorrupted words belong to $L(Obs_{opr}(G))$. Steps~4--10 guarantee that, in $Obs_{opr}(G)$, events in $E_+$ lead to the same states of the corresponding events in $E_{ins}$. Steps~11--15 guarantee that, in $Obs_{opr}(G)$, events in $E_-$ lead to self-loops. The above steps guarantee that all the stealthy words $w \in W_s$ belong to $L(Obs_{opr}(G))$.

Finally, Steps~16--22 impose that, if after executing Steps~1--15, a certain event in $E_a$ is not already enabled at a certain state of $Obs_{opr}(G)$, then such an event is enabled at such a state and leads to state $b_{\emptyset}$, where no other event may be executed. These steps ensure that all the exposing words $w \in W_e$ belong to $L(Obs_{opr}(G))$.

(b) We prove this by induction on the length of $w$. If $w=\varepsilon$, the result holds being $\hat P(w)=\varepsilon$.

Consider now a word $w \in L(Obs_{opr}(G))$ with length greater than one. Assume $w \in W_s$, and let $w=w'e_a$. Assume that the result holds for a generic $w'\in W_s$. By definition of operator mask, if $e_a \in E_o \cup E_+$, then $\widehat{P}(w) = \widehat{P}(w')e \in \widehat{P}(w')\{e, \varepsilon\}$; otherwise $\widehat{P}(w) = \widehat{P}(w') \in \widehat{P}(w')\{e, \varepsilon\}$. Thus $\delta_{opr}^*(b_0,w)=\delta_{opr}(\delta_{opr}^*(b_0,w'),e_a)$. Then, $\delta_{obs}^*(b_0,\widehat{P}(w))\!=\!\delta_{obs}(\delta_{obs}^*(b_0$, $\widehat{P}(w')),e)$ if $e_a \in E_o \cup E_+$, and
$\delta_{obs}^*(b_0,\widehat{P}(w))=\delta_{obs}(\delta_{obs}^*(b_0,\widehat{P}(w')),\varepsilon)$ if $e_a \in E_-$.

According to Algorithm~\ref{alg2} the transition function of $Obs_{opr}(G)$ starting for a generic state $b\in B$ is defined in the same way in case of $e$ and $e_+$ (Steps~6 and 7), while it corresponds to a self-loop in the case of $e_- \in E_-$ (Step~13). As a result, we can conclude that $\delta_{opr}^*(b_0,w)=\delta_{obs}^*(b_0,\widehat{P}(w))$.

Finally, the last claim in (b) follows from the fact that, if $w \in W_e$, according to Algorithm~\ref{alg2}, all the missing transitions end up in the new state $b_\emptyset$, thus $\delta_{opr}^*(b_0,w)=b_{\emptyset}$.~\hfill $\Box$

Given a plant \emph{G} with set of states \emph{X}, the observer of the plant $Obs(G)$ can be constructed in $2^{|X|}$ steps. According to Algorithm~\ref{alg2}, the operator observer $Obs_{opr}(G)$ has at most $2^{|X|}+1$ states; thus the complexity of constructing $Obs_{att}(G)$ is $O(2^{|X|})$.

\section{Unbounded and $n$-bounded joint estimators}\label{Attackstructure}

In this section we define a particular DFA, called \emph{joint estimator}, which is defined on alphabet $E_a$ and contains all attack words that can be generated by the plant.

Here we distinguish two different cases. In the first case, the attack function belongs to ${\mathcal F}_\infty$. We call \emph{unbounded joint estimator} the corresponding DFA, denoted as $A_\infty$. In the second case, the attack function belongs to ${\mathcal F}_n$ for a given $n \in \mathbb{N}$. We call \emph{$n$-bounded joint estimator} the corresponding DFA and denote it as $A_n$.

\subsection{Unbounded joint estimator}\label{Unboundedattackstructure}

Let us first formalize the definition of $A_{\infty}$.

\begin{defin} The \emph{unbounded joint estimator} $A_\infty=(R,E_a,\delta_a,r_0)$ w.r.t. \emph{G} and $E_{com}$ is defined as $A_\infty= Obs_{att}(G) \parallel Obs_{opr}(G)$.~\hfill $\diamond$ \label{def12} \end{defin}

\begin{figure}[htbp]
  \centering
  \includegraphics[width=3.2in]{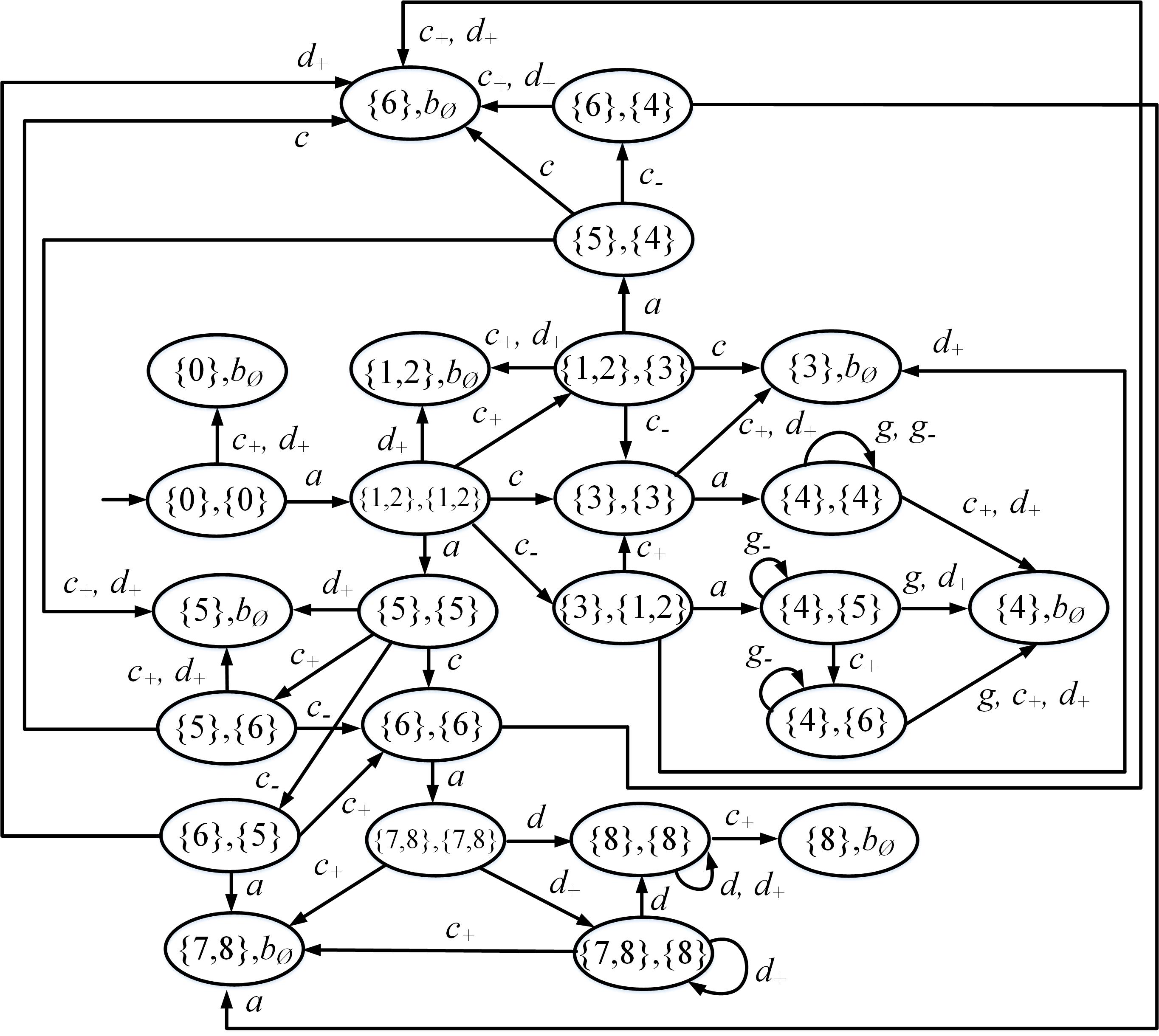}\\
  \caption{Unbounded joint estimator $A_\infty$ in Example~\ref{exa4}.}
  \label{fig9}
\end{figure}

\begin{example} Consider again the plant $G$ in Fig.~\ref{fig2} whose attacker observer and operator observer are visualized in Figs.~\ref{fig7}(a) and (b), respectively. The unbounded joint estimator $A_\infty$ built according to Definition~\ref{def12} is shown in Fig.~\ref{fig9}.
\label{exa4} \end{example}

By inspecting the unbounded joint estimator $A_\infty$ in Fig.~\ref{fig9}, once event $a$ occurs on the plant, the attacker executes event $a$ on $A_{\infty}$ starting from the initial state $(\{0\},\{0\})$. Thus state $(\{1,2\},\{1,2\})$ is reached. Now, the attacker may wait for a new event occurring on the plant, $a$ or $c$ in this case. Alternatively, the attacker may insert an event $c$ or $d$ in the operator observation, which correspond to execute $c_{+}$ or $d_{+}$, respectively, in $A_{\infty}$. Then, the attacker may erase event $c$ in the operator observation, which corresponds to execute $c_{-}$ in $A_{\infty}$, and so on.

We notice that, any word $w\in E_a^*$ generated by the unbounded joint estimator allows us to argue the following three information: (1) which is the observation $s\in E_o^*$ actually produced by the system; (2) how the attacker corrupted it (which events it has inserted and/or erased); (3) which is the word $s'\in E_o^*$ observed by the operator. Consider as an example, the word $w = a c_- a c_+ \in E_a^*$ that leads to state $(\{4\},\{6\})$. This corresponds to the observation $s=a ca $ produced by the system. The attacker erases the observation $c$ after the first $a$ and inserts the dummy observation $c$ after the second $a$, resulting in the word $s'=\widehat P(w)=a a c$ observed by the operator.~\hfill $\diamond$

The following theorem provides a characterization of the unbounded joint estimator.

\begin{theorem} Let $G$ be a plant with set of observable events $E_o$ and observer $Obs(G) = (B,{E_o},{\delta _{obs}},b_0)$. Given attack alphabet $E_a$, with set of stealthy words $W_s$  and set of exposing words $W_e$, let $A_\infty=(R,E_a,\delta _a,r_0)$ be its unbounded joint estimator. It holds that:
\label{the1} \end{theorem}

\begin{enumerate}
\item[(a)] $L(A_\infty) = L({\mathcal F}_{\infty},G) \cap \left (W_s \cup W_e \right)$;

\item[(b)] $\forall s \in P(L(G))$, $\forall f_{\infty}\in {\mathcal F}_{\infty}$ with $w=f_\infty(s) \in E_a^*$:
\begin{enumerate}
\item[(i)]
if $w \in W_s$ then $\delta_a^*(r_0,w)=(b_a,\overline{b}_a)$ $\Longleftrightarrow$ \\$\delta_{obs}^*(b_0,s)=b_a$, $\delta_{obs}^*(b_0,\widehat{P}(w))=\overline{b}_a$;
\item[(ii)]
if $w \in W_e$ then $\delta_a^*(r_0,w)=(b_a,b_\emptyset)$ $\Longleftrightarrow$ \\$\delta_{obs}^*(b_0,s)=b_a$, $\delta_{obs}^*(b_0,\widehat{P}(w))$ is not defined.
\end{enumerate}

\end{enumerate}

\emph{Proof:} (a) Follows from Propositions~\ref{prop1} and \ref{prop2} and Definition~\ref{def12}. Indeed, by Proposition~\ref{prop1}, it holds that $L(Obs_{att}(G))=L({\mathcal F}_{\infty},G)$ and by Proposition~\ref{prop2}, it holds that $L(Obs_{opr}(G)) = W_s \cup W_e$. Since $A_\infty$ is defined as the concurrent composition of $Obs_{att}(G)$ and $ Obs_{opr}(G)$, having the same alphabet, its language is equal to the intersection of the languages of the two DFA.

(b) We first consider the case: $w \in W_s$.

\emph{(If)} Assume that $\delta_{obs}^*(b_0,s)=b_a$, $\delta_{obs}^*(b_0,\widehat{P}(w))=\overline{b}_a$. By Propositions~\ref{prop1} and \ref{prop2}, it holds that $\delta_{obs}^*(b_0,s)=\delta_{att}^*(b_0,w)$ and $\delta_{obs}^*(b_0,\widehat{P}(w))=\delta_{opr}^*(b_0,w)$, namely, $\delta_{att}^*(b_0,w)=b_a$ and $\delta_{opr}^*(b_0,w)=\overline{b}_a$. Since $A_\infty=Obs_{att}(G) \parallel Obs_{opr}(G)$, by definition of concurrent composition, it is $\delta_a^*(r_0,w)=(b_a,\overline{b}_a)$.

\emph{(Only if)} Assume that $\delta_a^*(r_0,w)=(b_a,\overline{b}_a)$. Since $A_\infty=Obs_{att}(G) \parallel Obs_{opr}(G)$, by definition of concurrent composition, it holds that $\delta_{att}^*(b_0,w)=b_a$ and $\delta_{opr}^*(b_0,w)=\overline{b}_a$. By Propositions~\ref{prop1} and \ref{prop2}, it is $\delta_{obs}^*(b_0,s)=\delta_{att}^*(b_0,w)$ and $\delta_{obs}^*(b_0,\widehat{P}(w))=\delta_{opr}^*(b_0,w)$, namely, $\delta_{obs}^*(b_0,s)=b_a$ and $\delta_{obs}^*(b_0,\widehat{P}(w))=\overline{b}_a$.

Let us finally consider the case $w \in W_e$. The proof follows from the definition of concurrent composition and the fact that a word $w$ in the operator observer yields state $b_\emptyset$ if and only if $\widehat{P}(w) \not\in P(L(G))$, i.e., $\delta_{obs}^*(b_0,\widehat{P}(w))$ is not defined.~\hfill $\Box$

In plain words, Theorem~\ref{the1} implies that the language of the joint estimator contains all words on alphabet $E_a$ that can be generated under attack and that are either stealthy or exposing. In addition the state $(b_a,\overline{b}_a)$ reached in the joint estimator by a stealthy word $w=f_{\infty}(s)$ describes the joint estimation composed by the correct observation $b_a=\mathcal{C}(s)$ that would have been computed by the operator without attack, and the corrupted observation $\overline{b}_a=\mathcal{C}(s') = \mathcal{C}(\widehat{P}(w))$ due to the attack. An exposing word $w=f_{\infty}(s)$ reaches a state $(b_a,b_\emptyset)$ where $b_a=\mathcal{C}(s)$ is the correct observation that would have been computed by the operator without attack.

Let us show how to select a harmful attack function on the basis of the unbounded joint estimator $A_\infty$.

\begin{propo}
Given a plant $G=(X,E,\delta,x_0)$ with set of compromised events $E_{com}$, let $\mathcal{R} \subseteq 2^X \times 2^X$ be the misleading relation, and $A_\infty=(R,E_a,\delta _a,r_0)$ be the unbounded joint estimator. An attack function $f_n$ is harmful iff $R \cap \mathcal{R} \neq \emptyset$.
\label{prop3} \end{propo}
\emph{Proof:} \emph{(If)} Assume that there exists a state $r=(b_a,\overline{b}_a)$ of the joint estimator $A$ such that $r \in R \cap \mathcal{R}$. Since $r \in R$, then there exists an attack word $w$ with $\delta_a(r_0,w)=r$. According to the definition of attack function (Definition~\ref{def4}), there exists an observation $s \in P(L(G))$ such that $w=f_n(s)$, where $P$ is the natural projection.

According to Theorem~\ref{the1}, if $w \in W_s$ ($W_s$ is the set of stealthy words), we have $r=(b_a,\overline{b}_a)$ such that $\delta_{obs}^*(b_0,s)=b_a$, and $\delta_{obs}^*(b_0,\widehat{P}(w))=\overline{b}_a$, where $\widehat{P}$ is the operator mask. Note that, if $w \in W_e$ ($W_e$ is the set of exposing words), then $\delta_a(r_0,w)=r=(b_a,b_\emptyset) \notin \mathcal{R}$, which leads to a contradiction. For this reason, we exclude such a case.

Since $r=(b_a,\overline{b}_a) \in \mathcal{R}$, then the observation $s$ can be corrupted into a word $s'=\widehat{P}(w)$ such that $(\mathcal{C}(s),\mathcal{C}(s')) \in \mathcal{R}$, where $\mathcal{C}(s) = \delta_{obs}^*(b_0,s)$, and $\mathcal{C}(s') = \delta_{obs}^*(b_0,s')$ ($\mathcal{C}(s)$ (resp., $\mathcal{C}(s')$) is the set of states consistent with observation $s$ (resp., $s'$)). According to Definition~\ref{def7}, we can conclude that $f_n$ is harmful.

\emph{(Only if)} Assume that an attack function $f_n$ is harmful. According to Definition~\ref{def7}, there exists an observation $s$ that can be corrupted into an observation $s'=\widehat{P}(w)$ such that $(\mathcal{C}(s),\mathcal{C}(\widehat{P}(w))) \in \mathcal{R}$, where $\mathcal{C}(s) = \delta_{obs}^*(b_0,s)$, $\mathcal{C}(\widehat{P}(w)) = \delta_{obs}^*(b_0,\widehat{P}(w))$, and $w= f_n(s) \in W_s$ (we exclude the case that $w \in W_e$ for the same reason that discussed in the above proof).

Since the joint estimator $A$ contains all the possible attacks, and according to Theorem~\ref{the1}, then there must exist a state $r=\delta_a(r_0,w)=(b_a,\overline{b}_a)$ such that $\delta_{obs}^*(b_0,s)=b_a$, and $\delta_{obs}^*(b_0,\widehat{P}(w))=\overline{b}_a$. Namely, $R \cap \mathcal{R} \neq \emptyset$.~\hfill $\Box$


\begin{example}
Recall the plant $G$ in Fig.~\ref{fig2} with the unbounded joint estimator $A_\infty$ depicted in Fig.~\ref{fig9}. Assume that the misleading relation is $\mathcal{R}=\{(\{5\},X) \ | \ X \subseteq \{6,7,8\}\}$.
\label{ExaFinal} \end{example}

Looking at $A_\infty$, if the plant generates the word $aa$, $A_\infty$ is in state $(\{5\},\{5\})$, then the attacker can insert a fake event $c_+$ such that state $(\{5\},\{6\}) \in \mathcal{R}$ is reached. If such a state is reached, the plant is in the critical state $\{5\}$, while the operator thinks that the plant is in the non-critical state $\{6\}$. Thus no protective actions are activated, and damages are caused.~\hfill $\diamond$

We conclude this subsection discussing the complexity of computing the unbounded joint estimator $A_\infty$.

Given a plant \emph{G} with set of states \emph{X}, both the attack observer and the operator observer are computed in $2^{|X|}$ steps. The unbounded joint estimator is defined as $A_\infty=Obs_{att}(G) \parallel Obs_{opr}(G)$; thus the complexity of constructing $A_\infty$ is $O(2^{|X|} \cdot 2^{|X|})$.

\subsection{Bounded joint estimator}\label{nboundedattackstructure}

The $n$-bounded joint estimator $A_n$ that describes attack functions in ${\mathcal F}_n$, can be easily obtained starting from $A_{\infty}$. To this aim, a particular DFA, called \emph{n-bounded attack automaton}, denoted as $G_n$, is introduced. Then $A_n$ is obtained as the concurrent composition of $A_{\infty}$ and $G_n$.

\begin{defin}\label{def14}
The \emph{n-bounded attack automaton} is a DFA: $G_n=(X,E_a,\delta,0)$, where $X=\{0,1,\ldots, n\}$ ($n \in \mathbb{N}$), and the transition function is defined as follows:
\end{defin}

\begin{equation}
\left\{\begin{array}{lcl}
   {\forall i \in X, \ \delta(i,e_a):=0} \ \text{if} \ e_a \in E_a \setminus E_+, \\
   {\forall i \in (X \backslash \{n\}), \ \delta(i,e_a):=i+1} \ \text{if} \ e_a \in E_+.
   \end{array}\right.
\end{equation}~\hfill $\diamond$

\begin{figure}[htbp]
  \centering
  \includegraphics[width=3in]{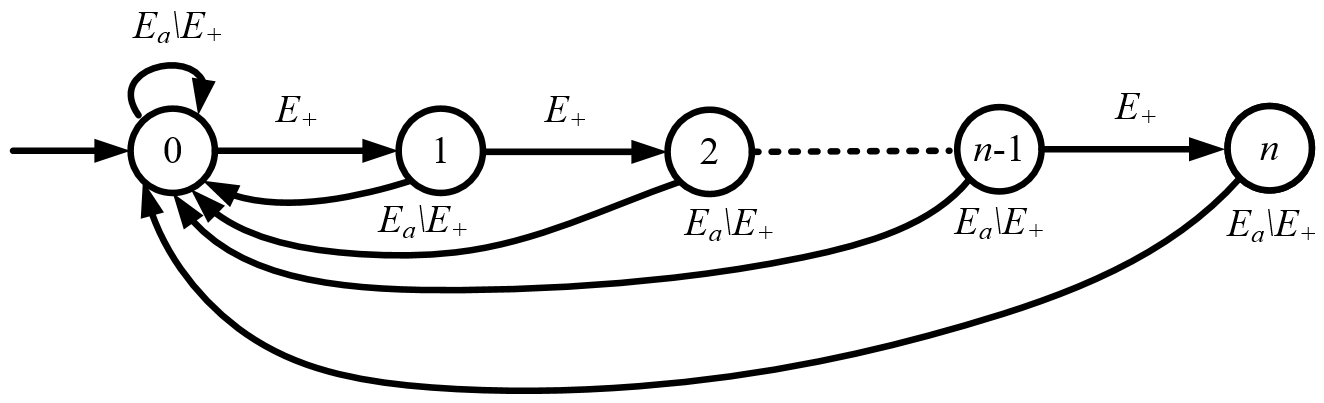}\\
  \caption{$n$-bounded attack automaton $G_n$}
  \label{fig8}
\end{figure}

Fig.~\ref{fig8} shows the $n$-bounded attack automaton $G_n$. As it can be seen, events in $E_a \setminus E_+$ are enabled at any state. On the contrary, events in $E_+$ are enabled provided that they have not been already executed $n$ times consecutively.

\begin{theorem} Let $G$ be a plant with attack alphabet $E_a$ and unbounded joint estimator $A_\infty=(R,E_a,\delta _a,r_0)$. Let $A_n= A_\infty \parallel G_n$, where $G_n$ is the $n$-bounded attack automaton. It holds that $L(A_n) =L(A_\infty) \setminus \{w \in L(A_\infty) \ | \ cons_{E_+}(w)>n \}$, where $cons_{E_+}(w)$ denotes the maximum number of consecutive events in $E_+$ contained in the word $w$.
\label{the2} \end{theorem}

\emph{Proof:} Follows from the fact that $A_n$ is defined as $A_n= A_\infty \parallel G_n$ and $G_n$ limits to $n$ the maximum number of consecutive events that the attacker can add to the operator observation.~\hfill $\Box$

\begin{figure}[htbp]
  \centering
  \includegraphics[width=2in]{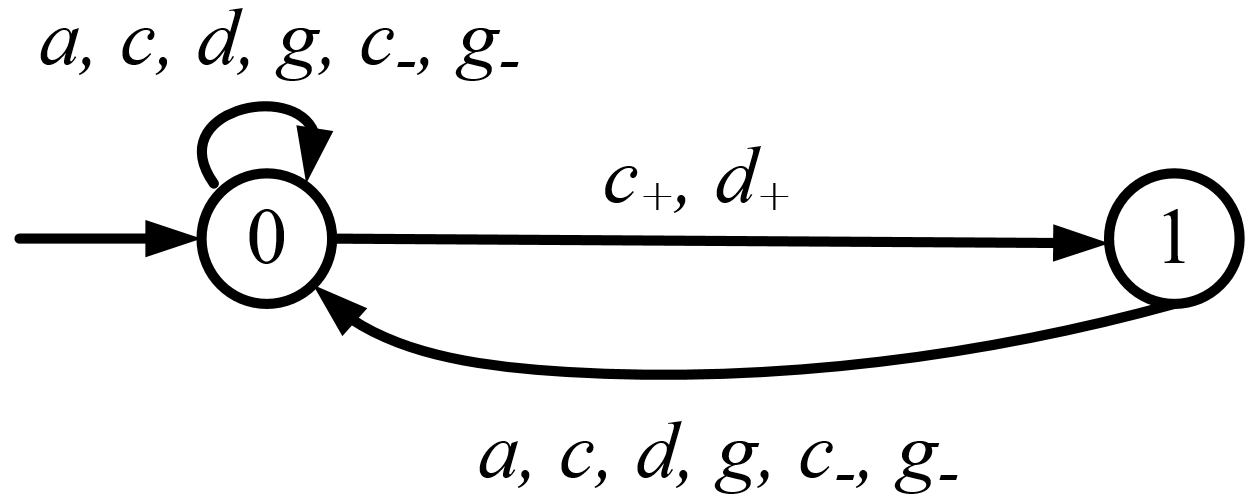}\\
  \caption{$1$-bounded attack automaton $G_1$ in Example~\ref{exa6}.}
  \label{fig10}
\end{figure}

\begin{figure}[htbp]
  \centering
  \includegraphics[width=3.2in]{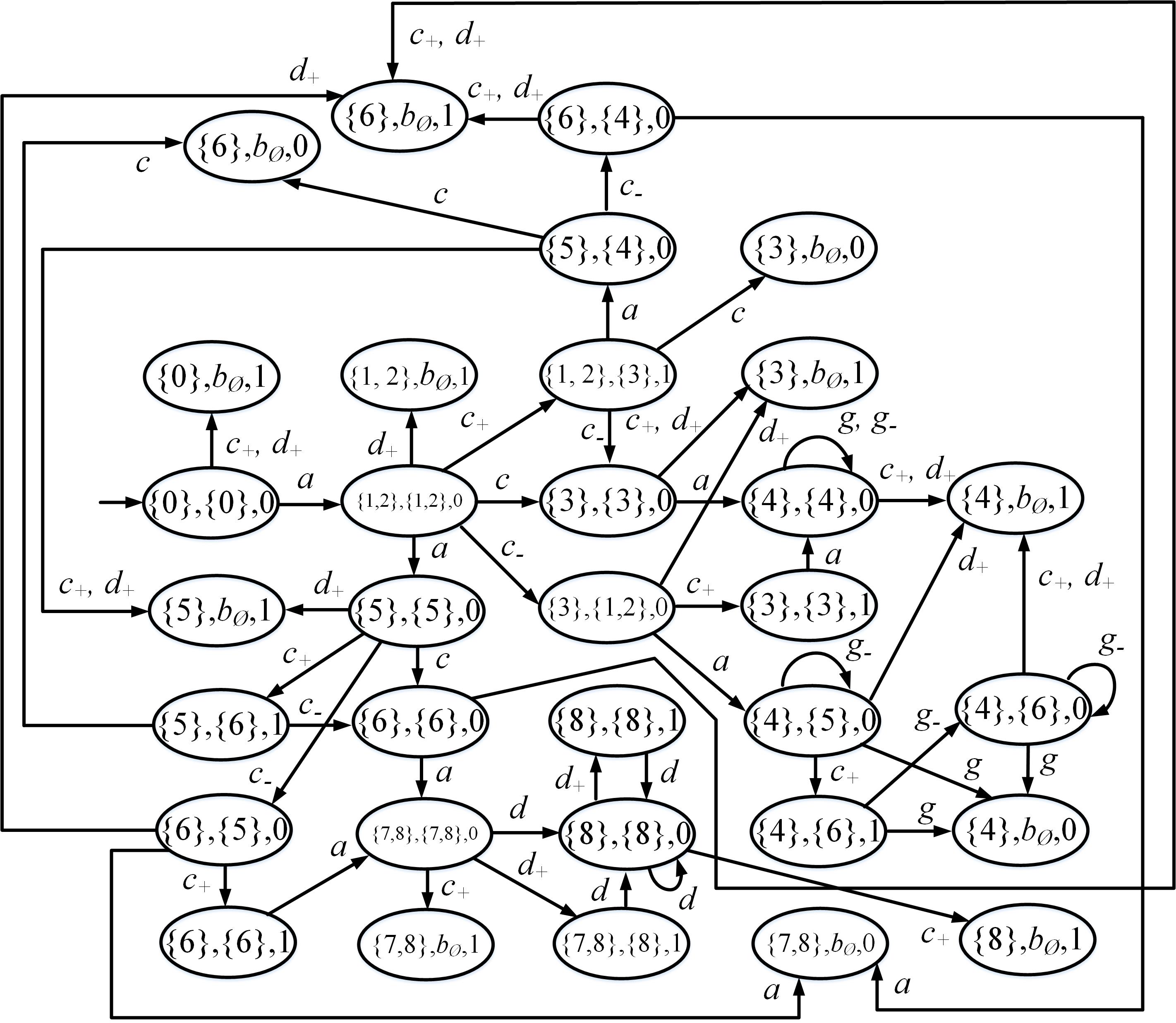}\\
  \caption{$1$-bounded joint estimator $A_1$ in Example~\ref{exa6}.}
  \label{fig11}
\end{figure}

\begin{example}
Consider the unbounded joint estimator $A_{\infty}$ in Example~\ref{exa4}. The $1$-bounded attack automaton $G_1$ and the $1$-bounded joint estimator $A_1=A_{\infty} \parallel G_1$ are depicted in Figs.~\ref{fig10} and~\ref{fig11}, respectively.
\label{exa6} \end{example}

\vspace{0.5cm}

The $1$-bounded joint estimator forces the attacker to insert at most one fake event between the occurrence of two observable events of the system.~\hfill $\diamond$

A result equivalent to Theorem~\ref{the1} holds for an $n$-bounded joint estimator.

Now we discuss the computational complexity of building the $n$-bounded joint estimator $A_n$. Given an integer value $n$, the $n$-bounded joint estimator is obtained by computing $A_n=A_\infty \parallel G_n$ where $G_n$ is the $n$-bounded attack automaton. Therefore, the complexity of computing $A_n$ is $O(2^{|X|}\cdot 2^{|X|} \cdot n)$.

\section{Conclusions and future work}\label{Conclusions}

In this paper we investigate the problem of state estimation under attack, for partially-observed discrete event systems. In more detail, an operator observes the system evolution with a natural projection, which {depends} on the sensors available on the system. The operator observation may be corrupted by an attacker. The corruption may be done by erasing some events that have occurred and/or inserting some events that have not actually occurred. It is possible to impose an upper bound on the number $n$ of consecutive observations that can be added by the attacker within the occurrence of two observable events in the plant. We show how to construct a joint estimator that contains all the possible attacks that can be implemented during the system evolution and that allows to establish if an attack function is harmful w.r.t. a given misleading relation.

Our future lines of research in this framework will follow several directions. On the one hand, we will look for a way to select stealthy and harmful attacks on the basis of the joint estimator or establish if such attacks can be thwarted. On the other hand, we will try to characterize and solve the same problem using Petri nets to understand if some advantages in terms of computational complexity can be obtained and if efficient solutions can also be computed for unbounded systems.

\bibliographystyle{IEEEtran}
\bibliography{IEEEabrv,IEEEexample}

\EOD

\end{document}